\documentclass[pra,showkeys,showpacs,byrevtex,twocolumn]{revtex4}%
\pdfoutput=1
\usepackage{amsfonts}
\usepackage{amsmath}
\usepackage{amssymb}
\usepackage{graphicx}%
\providecommand{\U}[1]{\protect\rule{.1in}{.1in}}

\newtheorem{theorem}{Theorem}

\newtheorem{corollary}{Corollary}

\newtheorem{lemma}{Lemma}

\newenvironment{proof}[1][Proof]{\noindent\textbf{#1.} }{\ \rule{0.5em}{0.5em}}
\begin{document}
\preprint{ }
\title[ ]{Can Classical Noise Enhance Quantum Transmission?}
\author{Mark M. Wilde}
\email{mark.wilde@usc.edu}
\affiliation{Communication Sciences Institute, Ming Hsieh Department of Electrical
Engineering, University of Southern California, Los Angeles, California USA 90089}
\affiliation{Hearne Institute for Theoretical Physics, Department of Physics and Astronomy,
Louisiana State University, Baton Rouge, Louisiana 70803}

\begin{abstract}
A modified quantum teleportation protocol broadens the scope of the classical
forbidden-interval theorems for stochastic resonance. The fidelity measures
performance of quantum communication. The sender encodes the two classical
bits for quantum teleportation as weak bipolar subthreshold signals and sends
them over a noisy classical channel. Two forbidden-interval theorems provide a
necessary and sufficient condition for the occurrence of the nonmonotone
stochastic resonance effect in the fidelity of quantum teleportation. The
condition is that the noise mean must fall outside a forbidden interval
related to the detection threshold and signal value. An optimal amount of
classical noise benefits quantum communication when the sender transmits weak
signals, the receiver detects with a high threshold, and the noise mean lies
outside the forbidden interval. Theorems and simulations demonstrate that both
finite-variance and infinite-variance noise benefit the fidelity of quantum teleportation.

\end{abstract}
\volumeyear{2008}
\volumeyear{2008}
\volumenumber{ }
\issuenumber{ }
\eid{ }
\date{\today}
\received[Received text]{}

\revised[Revised text]{}

\accepted[Accepted text]{}

\published[Published text]{}

\startpage{1}
\endpage{10}
\maketitle

\section{Introduction}

\begin{figure*}[tbp] \centering
\begin{tabular}
[c]{cc}%
{\parbox[b]{3.6236in}{\begin{center}
\includegraphics[
natheight=5.666200in,
natwidth=9.239700in,
height=1.9346in,
width=3.6236in
]%
{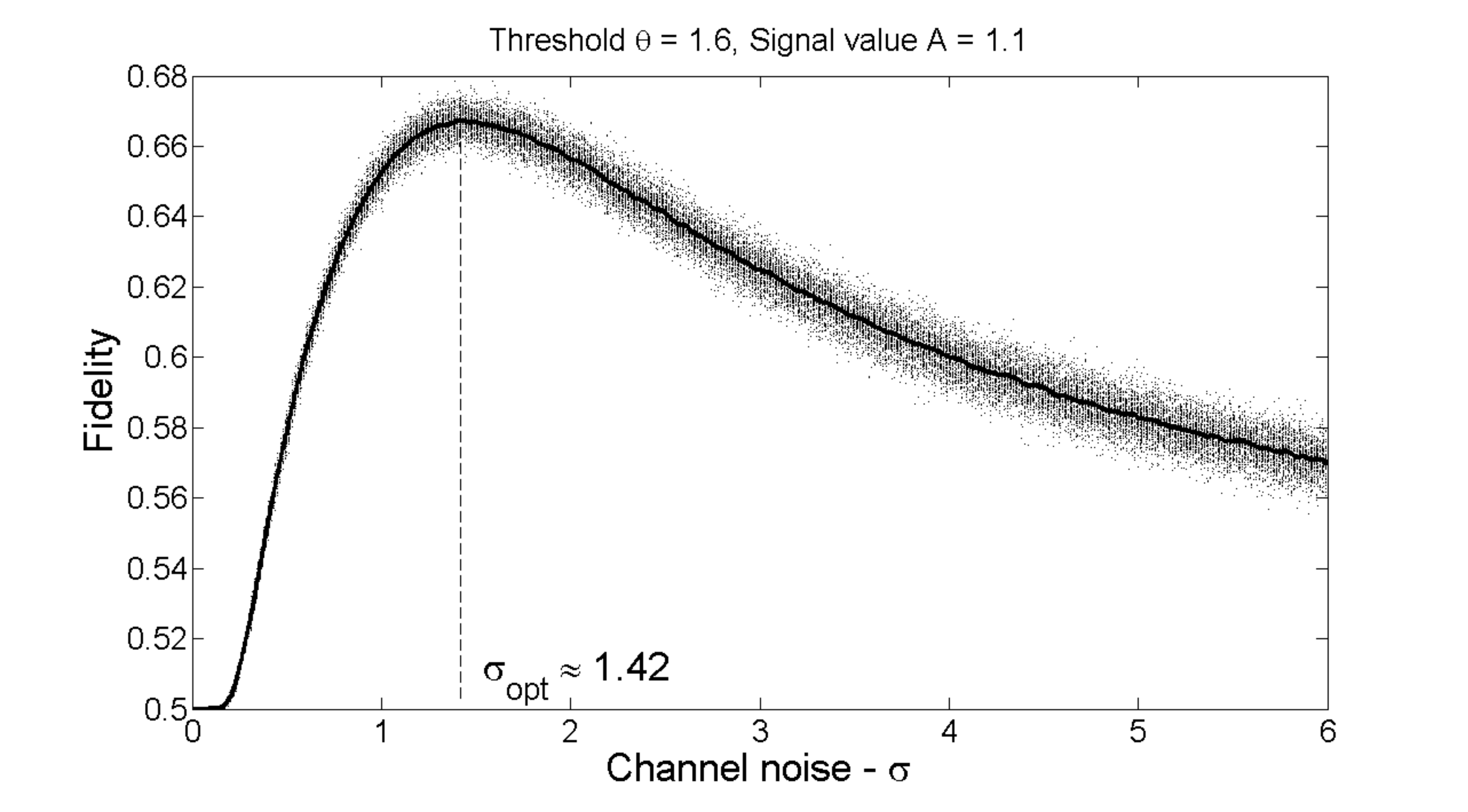}%
\\
(a)
\end{center}}}%
&
{\parbox[b]{3.6236in}{\begin{center}
\includegraphics[
natheight=5.066900in,
natwidth=8.460500in,
height=1.9346in,
width=3.6236in
]%
{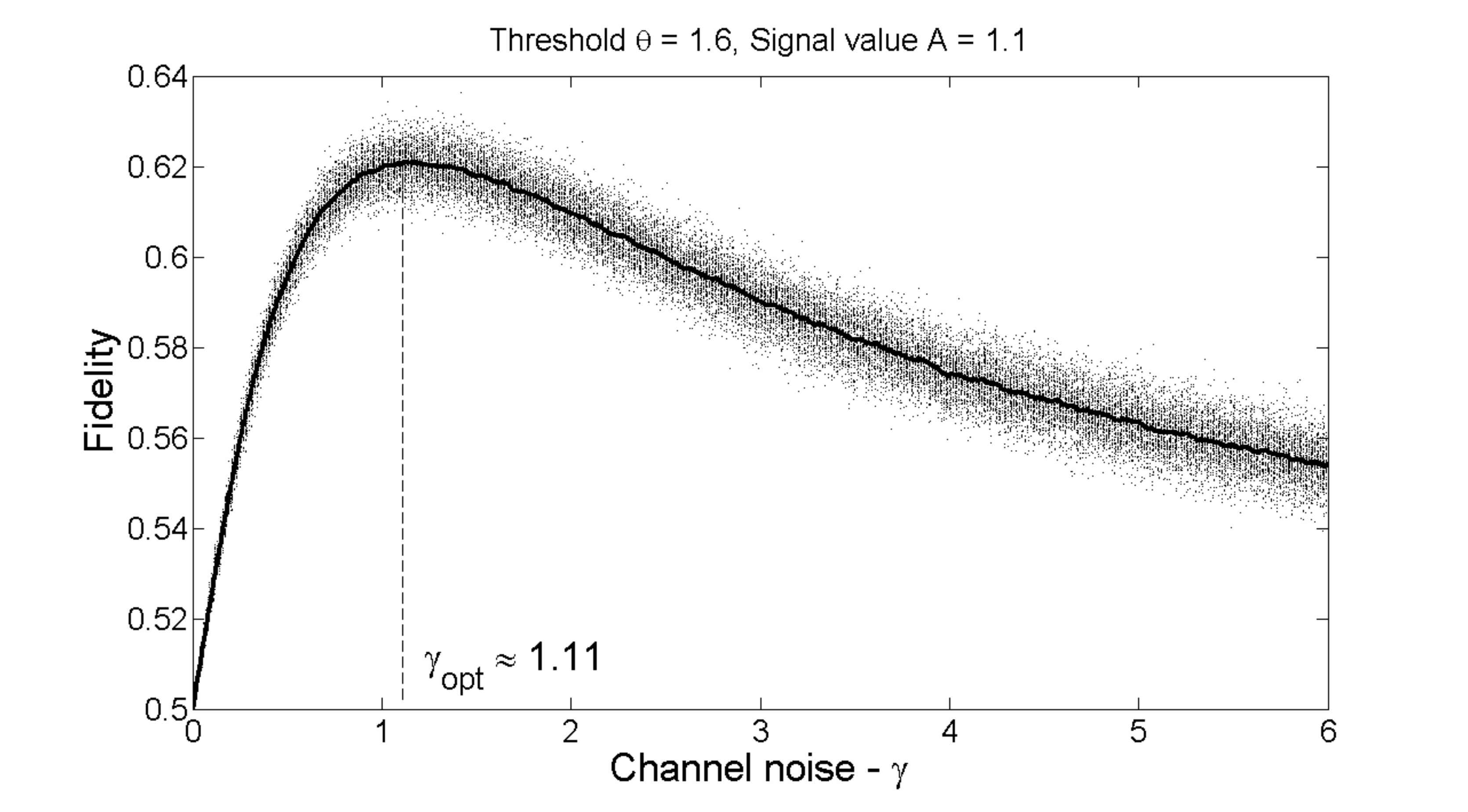}%
\\
(b)
\end{center}}}%
\end{tabular}%
\caption{Stochastic resonance in the modified teleportation protocol.
Alice possesses the state $(\left\vert0 \right\rangle+ \left\vert
1 \right\rangle)/\sqrt{2}$
and wishes to teleport it to Bob. The graphs show
the smoothed teleportation fidelity (thick line) and min-max deviation (dotted lines) as a function of (a) the variance of
classical Gaussian noise and a function of (b) the dispersion of
classical Cauchy noise for 100 simulation runs.
Alice encodes bipolar
signals with amplitude $A=1.1$ and Bob decodes with threshold $\theta
=1.6$. Each run generated 10 000 input-output
signal pairs to estimate the fidelity of teleportation.
Graph (a) is a simulation instance of the if-part
of Theorem~\ref{thm:finite}
with finite-variance Gaussian noise. The noise mean $\mu=0$
and lies outside the forbidden interval $(.5,2.7)$.
The average teleportation fidelity exceeds the
classical limit of 2/3 \cite{PhysRevLett.72.797} with $F=.6682$ for a
noise standard deviation $\sigma_{\text{opt}}=1.42$.
Graph (b) is a simulation instance of the if-part of
Theorem~\ref{thm:infinite}
with infinite-variance Cauchy noise. The noise location
$a=0$ and lies outside the forbidden interval $(.5,2.7)$. The average teleportation fidelity does not exceed the
classical limit of 2/3 with $F=.6213$ for a
noise dispersion $\gamma_{\text{opt}}=1.11$.}%
\label{fig:simulation}%
\end{figure*}

Noise can sometimes benefit the detection of weak signals
\cite{phystoday1996bulsara}. Researchers have dubbed this couterintuitive
phenomenon as the \textit{stochastic resonance} (SR) effect
\cite{revmod1998gamma}. A typical performance curve displays that performance
is poor for low noise values, increases to a maximal value for some optimal
noise, and tapers down again when too much noise is present
(Figure~\ref{fig:simulation}\ displays such performance curves).

The SR noise benefit occurs in a diverse range of systems from neurons
\cite{PhysRevE.53.3958}\ to superconducting quantum interference devices
\cite{squid1995}\ to crayfish \cite{nature1995wiesenfeld}. The SR\ noise
benefit also occurs in the quantum regime with unique quantum effects such as
squeezed light \cite{qcmc2006wilde,WK08}, tunneling \cite{PhysRevLett.72.1947}%
, quantum jumps in a micromaser \cite{PhysRevLett.80.3932}, electron shelving
\cite{PhysRevA.62.052111}, and entanglement \cite{PhysRevLett.88.197901}. All
the aforementioned classical and quantum scenarios for SR\ involve the
detection of weak signals.

The ingredients for a noise benefit are weak signals, a nonlinear detection
scheme, and a source of noise energy. Noise energy does not benefit
communication in linear systems because amplification only increases the noise
in the signal. But small amounts of noise energy can be beneficial in a simple
nonlinear threshold detection scheme. It can boost the signal above a
threshold when it otherwise would be undetectable. The noise benefit occurs in
most nonlinear systems because they act as threshold systems at some level.

The classical forbidden-interval theorems \cite{nn2003kosko,pre2004kosko}
apply to a simple threshold system. The theorems give necessary and sufficient
conditions for an SR noise benefit in a memoryless threshold neuron. The
communication model for the theorems has a simple form in terms of a threshold
function with threshold $T$ and subthreshold bipolar signals with values $-A$
and $A$ where $-A<0<A<T$. The forbidden-interval condition is that an
SR\ noise benefit occurs if and only if the noise mean does not lie in the
interval $\left(  T-A,T+A\right)  $. The significance of the theorems is that
the forbidden-interval condition implies that the SR noise benefit occurs in a
memoryless threshold neuron for \textit{any} finite-variance noise or
infinite-variance alpha-stable noise \cite{book1995nikias}.

I broaden the scope of the classical forbidden-interval theorems by
constructing a modified teleportation protocol in which classical noise
enhances the fidelity of quantum teleportation (Figure~\ref{fig:mod-tele}).
This phenomenon is an SR\ noise benefit because the enhancement occurs for
some optimal non-zero classical noise level. The original quantum
teleportation protocol uses one ebit of shared entanglement and two noiseless
feedforward classical bits to transmit one qubit \cite{PhysRevLett.70.1895}.
Later work considers a noisy teleportation protocol that sends quantum
information with a noisy classical channel \cite{DHW05RI}. I consider a
similar teleportation model where the entanglement is noiseless and the
classical communication is over a noisy classical channel. But in this
protocol, the transmitter Alice encodes the two classical bits as two weak,
subthreshold, bipolar classical signals and sends them over the noisy channel.
A receiver Bob then thresholds to determine the two classical bits Alice sent.
This modified teleportation protocol then leads to an SR noise benefit for the
fidelity of quantum communication.

The fidelity for the modified teleportation protocol qualitatively behaves
similarly to the mutual information measure for the classical SR noise benefit
in neurons in \cite{nn2003kosko,pre2004kosko}. The similarity is qualitative
because the fidelity displays the full inverted--U signature of the SR noise
benefit given satisfaction of the forbidden-interval condition. But the
fidelity measures quantum communication performance while the mutual
information measures classical communication performance.
Theorems~\ref{thm:finite} and \ref{thm:infinite} have the same
forbidden-interval condition but now apply to the fidelity measure.

Two forbidden-interval theorems---Theorems~\ref{thm:finite} and
\ref{thm:infinite}---give necessary and sufficient conditions for the
SR\ noise benefit in quantum transmission. The first theorem holds for any
finite-variance noise and the second theorem holds for infinite-variance
alpha-stable noise. The proof strategy for Theorems~\ref{thm:finite} and
\ref{thm:infinite} is the same as the earlier strategy in
\cite{nn2003kosko,pre2004kosko}. The original proof strategy in
\cite{nn2003kosko,pre2004kosko} constructed all crucial arguments in terms of
detection probabilities. The simple and elegant expression for the fidelity in
Lemma~\ref{lem:fidelity}\ in terms of detection probabilities implies that the
same proof strategy is applicable. The proof strategy is to show that the
fidelity must increase from its minimum of 1/2 with the addition of classical
noise. This increase occurs if the fidelity approaches its minimum of 1/2 as
the variance or the dispersion of the noise decreases to zero. In
Section~\ref{sec:imperfect}, I show that the SR\ effect occurs even when the
entanglement resource is imperfect, i.e., if the sender and receiver share
noisy entanglement. Theorems~\ref{thm:finite} and \ref{thm:infinite} provide a
theoretical underpinning to explain why the SR effect occurs in the modified
teleportation protocol just as the original forbidden-interval theorems
explain why the SR effect occurs in a noisy threshold neuron.

Ting has previously considered the SR\ effect in quantum communication
\cite{ting99,ting00,PhysRevE.59.2801}. He specifically considered the response
of the coherent information and the fidelity to noise in several types of
Pauli channels. He found that the coherent information quantum information
measure does not exhibit a noise-enhanced SR\ effect, but the fidelity does
exhibit such an effect. The present work is similar to his because I consider
the fidelity of quantum communication as the measure of performance, but the
model under which the stochastic resonance effect occurs is significantly
different because I employ a modified teleportation protocol with subthreshold
classical signals, while he considered the effect of transmitting qubits over
noisy qubit channels.

\section{Model for Stochastic Resonance in Quantum
Teleportation\label{sec:model}}%

\begin{figure}
[ptb]
\begin{center}
\includegraphics[
natheight=5.266700in,
natwidth=11.326400in,
height=1.516in,
width=3.2491in
]%
{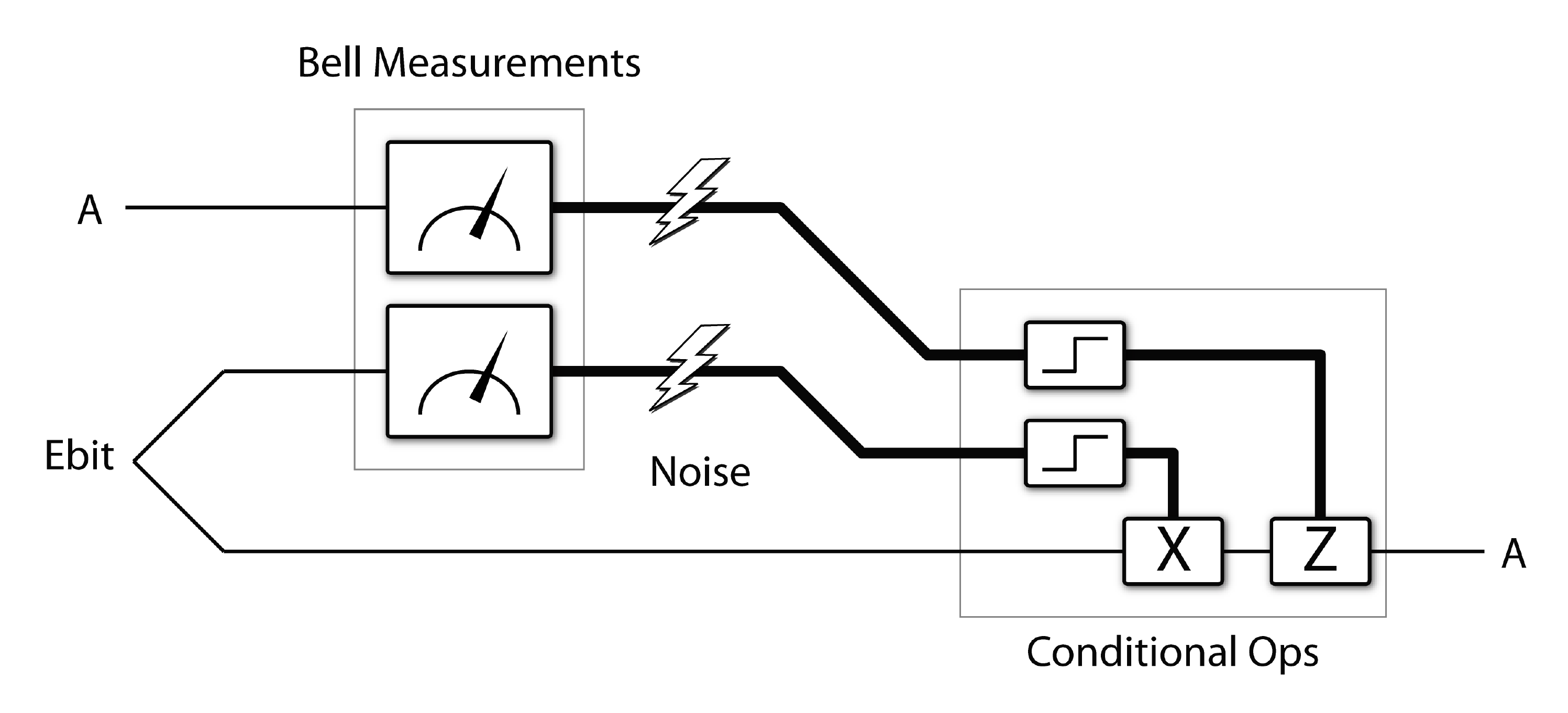}%
\caption{Modified quantum teleportation protocol with a noisy classical
channel and thresholding. Thick lines denote classical information and thin
lines denote quantum information. Alice wants to teleport quantum state $A$ to
Bob. Alice and Bob share an ebit. Alice receives two classical bits from a
Bell measurement of her state $A$ and her half of the shared ebit. Alice
encodes the two classical bits as weak bipolar signals and transmits them over
a noisy channel. Bob thresholds the signal he receives to retrieve two
classical bits. Bob then performs a conditional rotation of his state by
$\hat{X}$ or $\hat{Z}$ or both in the hope that he rotates his state to be
Alice's original state $A$.}%
\label{fig:mod-tele}%
\end{center}
\end{figure}
I first review the noiseless quantum teleportation protocol
\cite{PhysRevLett.70.1895} before presenting the modified teleportation
protocol. Suppose that Alice and Bob share one ebit---a maximally entangled
quantum state $\left\vert \Phi^{+}\right\rangle ^{AB}\equiv(\left\vert
0\right\rangle ^{A}\left\vert 0\right\rangle ^{B}+\left\vert 1\right\rangle
^{A}\left\vert 1\right\rangle ^{B})/\sqrt{2}$. Suppose Alice wants to transmit
a quantum bit $\left\vert \psi\right\rangle ^{A^{\prime}}=\alpha\left\vert
0\right\rangle ^{A^{\prime}}+\beta\left\vert 1\right\rangle ^{A^{\prime}}$ to
Bob. Define quantum state $\left\vert \phi\right\rangle ^{A^{\prime}AB}$ as
the joint state of system $A^{\prime}$ and ebit $\left\vert \Phi
^{+}\right\rangle ^{AB}$:%
\begin{equation}
\left\vert \phi\right\rangle ^{A^{\prime}AB}\equiv\left\vert \psi\right\rangle
^{A^{\prime}}\otimes\left\vert \Phi^{+}\right\rangle ^{AB}.
\label{eq:joint-state}%
\end{equation}
Alice can teleport state $A^{\prime}$ to Bob by performing a two-qubit Bell
measurement on qubit $A^{\prime}$ and on her half $A$ of the shared ebit.
Alice receives two classical bits $s_{1}s_{2}$ from the Bell measurement where
$\forall i\in\left\{  1,2\right\}  ,\ s_{i}\in\left\{  0,1\right\}  $. The
Bell measurement is probabilistic so that bits $s_{1}$ and $s_{2}$ are
realizations of two Bernoulli random variables $S_{1}$ and $S_{2}$
respectively. The following useful lemma gives the density of the two
classical bits $s_{1}s_{2}$ that Alice receives from the Bell measurement.
Lemma~\ref{lem:independent-equi} follows simply from the original
teleportation protocol \cite{PhysRevLett.70.1895}. The appendix gives the
proof of the following lemma and of all following lemmas and theorems.

\begin{lemma}
\label{lem:independent-equi}Random variables $S_{1}$ and $S_{2}$\ from the
Bell measurement are independent and identically distributed with equal
probability of being zero or one:%
\begin{equation}
P_{S_{i}}\left(  s_{i}\right)  =1/2\ \ \ \ \ \forall\ i\in\left\{
1,2\right\}  \ \ \forall\ s_{i}\in\left\{  0,1\right\}  .
\end{equation}

\end{lemma}

Alice transmits the two classical bits $s_{1}s_{2}$ over a noiseless classical
channel. Bob receives the two classical bits and performs a conditional
rotation $\hat{Z}^{s_{2}}\hat{X}^{s_{1}}$ on his half $B$ of the shared ebit.
$\hat{Z}$ and $\hat{X}$ are the Pauli operators \cite{book2000mi}. The
teleportation is a perfect success if Alice can perform a perfect Bell
measurement, if Alice sends two \textit{noiseless} classical bits to Bob, and
if Bob can perform the conditional unitaries without any small error in the
rotation. The state in Bob's lab $B$ is $\left\vert \psi\right\rangle
^{B}=\alpha\left\vert 0\right\rangle ^{B}+\beta\left\vert 1\right\rangle ^{B}$
when teleportation is perfect.

I now construct a modified teleportation protocol that uses
\textit{subthreshold} classical signals (Figure \ref{fig:mod-tele}). This
model leads to an SR\ noise benefit for the fidelity of teleportation. Suppose
Alice still performs a perfect two-qubit Bell measurement on her qubit
$\left\vert \psi\right\rangle ^{A^{\prime}}$ and her half $A$ of the shared
ebit. Alice receives two random classical bits $s_{1}$ and $s_{2}$ from the
Bell measurement. Let $S$ be a Bernoulli random variable with equal
probability for outcome zero or outcome one. Bits $s_{1}$ and $s_{2}$ are
independent realizations of random variable $S$ by
Lemma~\ref{lem:independent-equi}. Suppose Alice cannot transmit noiseless
classical bits and must instead use a continuous additive noisy classical
channel for transmission\ \cite{book1991cover}. Suppose further that Alice
sends two weak, bipolar, subthreshold signals over the additive noisy
classical channel. She encodes the random bits with the map $\left(
-1\right)  ^{S+1}A$ so that signal value $-A$ corresponds to `0' and signal
value $A$ corresponds to `1'. The signals are weak in the sense that they are
\textit{subthreshold}---the threshold $\theta$ is larger than the signal
values:\ $-A<0<A<\theta$. The additive noisy channel corrupts the two
classical signals by adding a random noise $N$. Suppose the noise $N$ for two
uses of the channel is independent and identically distributed. The noise $N$
and random variable $S$ are independent because the noise $N$ plays no role in
the Bell measurement. The two signals Bob receives from both uses of the
channel are independent realizations of random variable $\left(  -1\right)
^{S+1}A+N$. Suppose Bob detects the classical signals by thresholding with a
threshold~$\theta$. He counts a `1' if the signal he receives is greater than
$\theta$ and counts a `0' if the signal is less than~$\theta$. Let $y_{1}$ and
$y_{2}$ be the two bits from Bob's detection. Both bits are independent
realizations of a random variable $Y$ where%
\begin{equation}
Y=u\left(  \left(  -1\right)  ^{S+1}A+N-\theta\right)  ,
\end{equation}
and $u$ is the unit step or Heaviside function. The quantum state Bob
possesses after Alice performs the Bell measurement is $\left\vert \psi
_{s_{1}s_{2}}\right\rangle ^{B}\equiv\hat{Z}^{s_{2}}\hat{X}^{s_{1}}\left\vert
\psi\right\rangle ^{B}$. Bob does not have knowledge of bits $s_{1}$ and
$s_{2}$ so he cannot rotate his state to be the same as Alice's original qubit
$\left\vert \psi\right\rangle ^{A}$ with probability one. He can perform a
rotation of his state based only on bits $y_{1}$ and $y_{2}$. So Bob performs
a conditional rotation $\hat{Z}^{y_{2}}\hat{X}^{y_{1}}$ in an attempt to
rotate the state $\left\vert \psi_{s_{1}s_{2}}\right\rangle ^{B}$ to state
$\left\vert \psi\right\rangle ^{B}$. Suppose Bob performs a noiseless Pauli
$\hat{Z}$, $\hat{X}$, or $\hat{Z}\hat{X}$ gate when he performs the
conditional rotation. His resulting state is $\left\vert \psi_{y_{1}y_{2}%
s_{1}s_{2}}\right\rangle ^{B}\equiv\hat{Z}^{y_{2}}\hat{X}^{y_{1}}\left\vert
\psi_{s_{1}s_{2}}\right\rangle ^{B}$. He does not apply the proper
rotation\ if $y_{1}y_{2}\neq s_{1}s_{2}$. Thus Bob's state is a mixture
$\rho_{B}$ equal to the following matrix:%
\begin{equation}
\sum_{\substack{y_{1},y_{2},\\s_{1},s_{2}=0}}^{1}p_{Y_{1},Y_{2},S_{1},S_{2}%
}\left(  y_{1},y_{2},s_{1},s_{2}\right)  \left\vert \psi_{y_{1}y_{2}s_{1}%
s_{2}}\right\rangle \left\langle \psi_{y_{1}y_{2}s_{1}s_{2}}\right\vert ,
\label{eq:bobs-state}%
\end{equation}
where $p_{Y_{1},Y_{2},S_{1},S_{2}}$ is the joint probability distribution of
random variables $Y_{1}$, $Y_{2}$, $S_{1}$, and $S_{2}$ (we evaluate it later
on). The modified teleportation protocol leads to noisy quantum communication
because Bob's final state is the noisy mixed state above. Alice cannot
teleport her state perfectly to Bob in the modified teleportation protocol
with Alice encoding with subthreshold signals and Bob detecting with a
threshold system.

The fidelity measure quantifies the quality of Alice and Bob's quantum
communication \cite{book2000mi}. The fidelity $F$ is as follows%
\begin{equation}
F\equiv\left\langle \psi\right\vert \rho_{B}\left\vert \psi\right\rangle ,
\end{equation}
where $\left\vert \psi\right\rangle $ is Alice's original state $\left\vert
\psi\right\rangle ^{A^{\prime}}$ and $\rho_{B}$ is Bob's mixed state from
(\ref{eq:bobs-state}). Several example values of the fidelity eludicate some
meaning behind this measure of quantum communication. The fidelity $F=1$ if
and only if Alice's state is the same as Bob's state. $F=0$ if and only if
Alice and Bob's states are orthogonal. Suppose Bob ignores the classical
information Alice sends in the teleportation protocol, randomly chooses a
rotation, and does not record which rotation he performs. Then his state is
maximally mixed with density matrix $\rho_{B}=I/2$. So Alice and Bob's
fidelity $F=1/2$ if $\rho_{B}=I/2$. Alice and Bob can obtain a fidelity of
teleportation equal to $2/3$ even when they don't share entanglement and use
only noiseless classical communication \cite{PhysRevLett.72.797}.

The fidelity for the modified teleportation protocol admits a simple
mathematical form in terms of four quantities: $q_{X}$, $q_{Z}$, $q_{XZ}$, and
$P$. Define $q_{X}$, $q_{Z}$, $q_{XZ}$ as%
\begin{align}
q_{Z}  &  \equiv\left\vert \left\langle \psi\right\vert \hat{Z}\left\vert
\psi\right\rangle \right\vert ^{2},\label{eq:q_Z}\\
q_{X}  &  \equiv\left\vert \left\langle \psi\right\vert \hat{X}\left\vert
\psi\right\rangle \right\vert ^{2},\\
q_{XZ}  &  \equiv\left\vert \left\langle \psi\right\vert \hat{X}\hat
{Z}\left\vert \psi\right\rangle \right\vert ^{2}. \label{eq:q_XZ}%
\end{align}
The quantities $q_{X}$, $q_{Z}$, and $q_{XZ}$ depend on the probability
amplitudes $\alpha$, $\beta$ of Alice's state $\left\vert \psi\right\rangle
^{A^{\prime}}$. The quantities $q_{X}$, $q_{Z}$, and $q_{XZ}$ are nonnegative
and convex so that $q_{X}+q_{Z}+q_{XZ}=1$. Define the nonnegative quantity $P$
as the difference of conditional probabilities:%
\begin{equation}
P\equiv p_{Y|S}\left(  0|0\right)  -p_{Y|S}\left(  0|1\right)  =p_{Y|S}\left(
1|1\right)  -p_{Y|S}\left(  1|0\right)  .
\end{equation}
The proof of the nonnegativity of $P$ and the equality of the above
conditional probability differences is in the proof of
Lemma~\ref{lem:fidelity}. Note that equality of $p_{Y|S}\left(  0|0\right)
-p_{Y|S}\left(  0|1\right)  $ and $p_{Y|S}\left(  1|1\right)  -p_{Y|S}\left(
1|0\right)  $ holds because the classical signals in the model are
subthreshold. As a simple example of this equality, note that $p_{Y|S}\left(
1|1\right)  =0$, $p_{Y|S}\left(  1|0\right)  =0$, $p_{Y|S}\left(  0|0\right)
=1$, and $p_{Y|S}\left(  0|1\right)  $ if there is no noise on the classical
channel. Consider that Lemma~\ref{lem:fidelity}\ gives the simple mathematical
expression for the fidelity $F$.

\begin{lemma}
\label{lem:fidelity}The fidelity $F$ between Alice's initial quantum state
$\left\vert \psi\right\rangle ^{A^{\prime}}$ and Bob's mixed state $\rho_{B}$
is%
\begin{equation}
F=\frac{1}{2}+\frac{P\left(  q_{X}+q_{Z}+q_{XZ}P\right)  }{2},
\label{eq:fidelity-final}%
\end{equation}
given the modified teleportation protocol.
\end{lemma}

The noisy classical channel affects only parameter $P$ and parameter $P$
varies between zero and one depending on how much noise is present in the
channel. The other parameters $q_{X}$, $q_{Z}$, and $q_{XZ}$ depend on the
quantum state $\left\vert \psi\right\rangle ^{A^{\prime}}$ that Alice wishes
to teleport---they depend on the probability amplitudes $\alpha$, $\beta$. The
noisy channel does not affect $q_{X}$, $q_{Z}$, and $q_{XZ}$ so that the
fidelity changes with the noisiness of the channel regardless of the quantum
state that Alice teleports.

The mutual information measure for classical SR\ has a more complicated
relationship with parameter $P$ than does the above fidelity measure
\cite{nn2003kosko,pre2004kosko}. It is elegant that the fidelity measure for
quantum communication has such a simple quadratic relation with parameter $P$
given the modified teleportation protocol.%

Corollary~\ref{lem:minimum-fidelity} relates the fidelity of teleportation to
the statistical relationship between random variables $S$ and $Y$. The
relationship follows by determining the quantity $P$ when random variables $S$
and $Y$ are statistically dependent, statistically independent, and when $S$
and $Y$ correlate perfectly. The relationship of the fidelity $F$ with random
variables $S$ and $Y$\ follows directly from the relationship of parameter $P$
with $S$ and $Y$ by using (\ref{eq:fidelity-final}).

\begin{corollary}
\label{lem:minimum-fidelity}The fidelity $F$ between Alice's initial quantum
state $\left\vert \psi\right\rangle ^{A^{\prime}}$ and Bob's mixed state
$\rho_{B}$ is minimum at $1/2$ given the modified teleportation protocol. The
fidelity $F$ obtains this minimum value if and only if random variable $Y$ is
independent of random variable $S$. The fidelity $F>1/2$ if $Y$ and $S$ are
statistically dependent. The fidelity $F$ is equal to its maximum of one when
detection is perfect.
\end{corollary}

Corollary~\ref{lem:minimum-fidelity} is useful because it provides both a
lower and upper bound for the fidelity of teleportation given the modified
teleportation protocol. It also gives the scenarios in which these lower and
upper bounds saturate. The fidelity cannot decrease below $1/2$ for any amount
of noise in the classical channel. This lower bound is a powerful way to
characterize the stochastic resonance effect in terms of the fidelity. The SR
noise benefit has a nonmonotone signature because the performance measure
decreases as the noise level decreases. So the fidelity should decrease to its
minimum of $1/2$\ when the noise variance or dispersion of the channel
decreases to zero. This statement is equivalent to saying that the fidelity
increases from its minimum value of $1/2$ as the noise variance or dispersion
of the channel increases:\ \textit{what goes down must come up}. The if-part
of the theorems employ the \textit{what goes down must come up} strategy to
show that the fidelity approaches its minimum of 1/2 when the noise vanishes
similar to the way that the mutual information approaches its minimum of zero
when the noise vanishes \cite{nn2003kosko}.
Corollary~\ref{lem:minimum-fidelity} is also useful because it gives the
situation in which the fidelity is equal to its maximum of one. This situation
provides a powerful way of determining when the SR noise benefit does not
occur. The SR noise benefit does not occur when the fidelity of teleportation
increases to its maximum value of one as the noise variance or dispersion
decreases to zero. The only-if part of the theorems show that the fidelity
approaches its maximum value of one as the noise vanishes similar to the way
that the mutual information approaches its maximum of one as the noise
vanishes \cite{pre2004kosko}. I employ these proof strategies involving the
lower and upper bounds from Corollary~\ref{lem:minimum-fidelity} in the proofs
of the main results:\ Theorems~\ref{thm:finite}~and~\ref{thm:infinite}.

\section{Forbidden-Interval Theorems for Quantum Teleportation}

\subsection{SR\ with Finite-Variance Noise}%

\begin{figure*}[tbp] \centering
\begin{tabular}
[c]{cc}%
{\parbox[b]{3.6236in}{\begin{center}
\includegraphics[
natheight=7.073300in,
natwidth=13.333700in,
height=1.9346in,
width=3.6236in
]%
{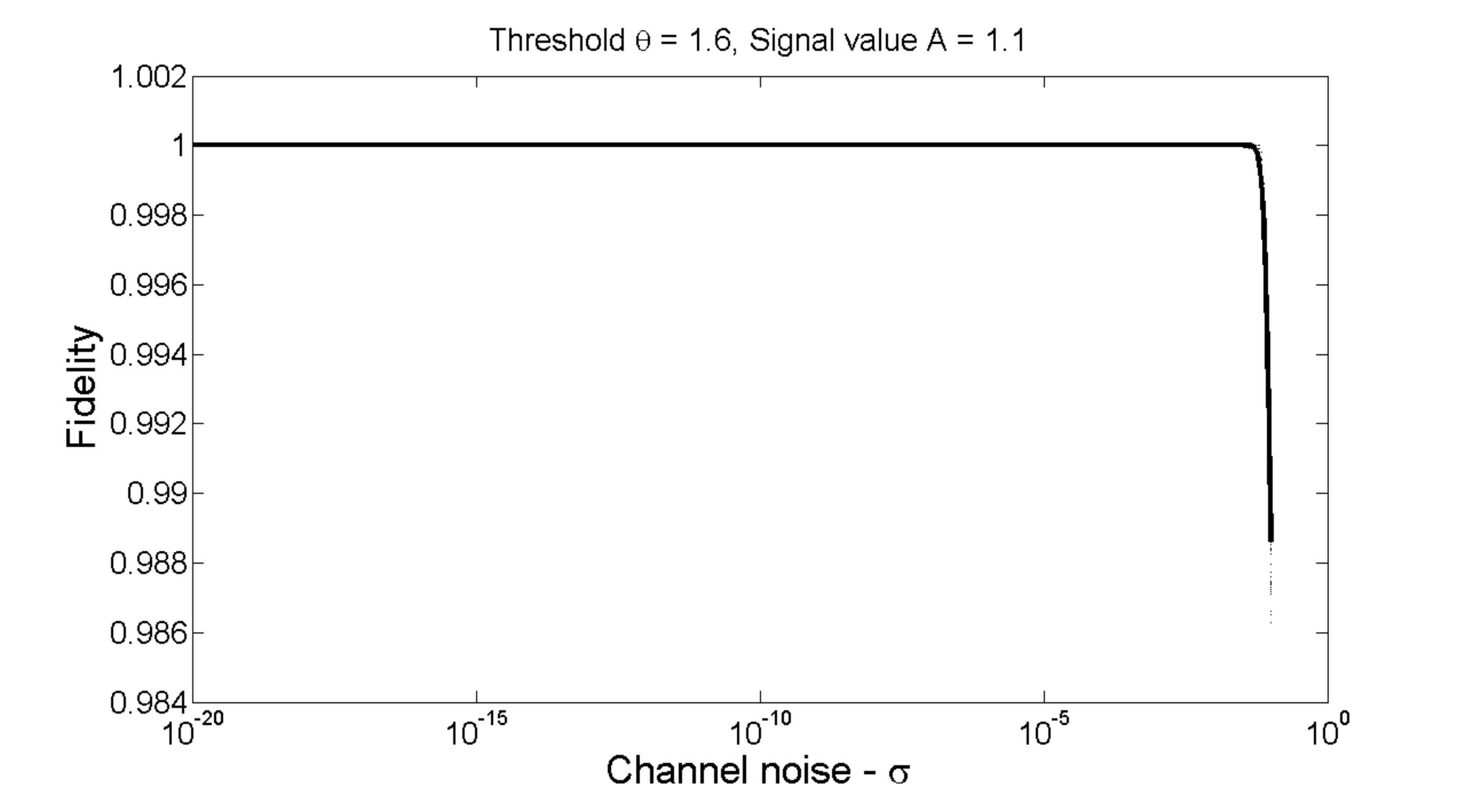}%
\\
(a)
\end{center}}}%
&
{\parbox[b]{3.6236in}{\begin{center}
\includegraphics[
natheight=7.073300in,
natwidth=13.333700in,
height=1.9346in,
width=3.6236in
]%
{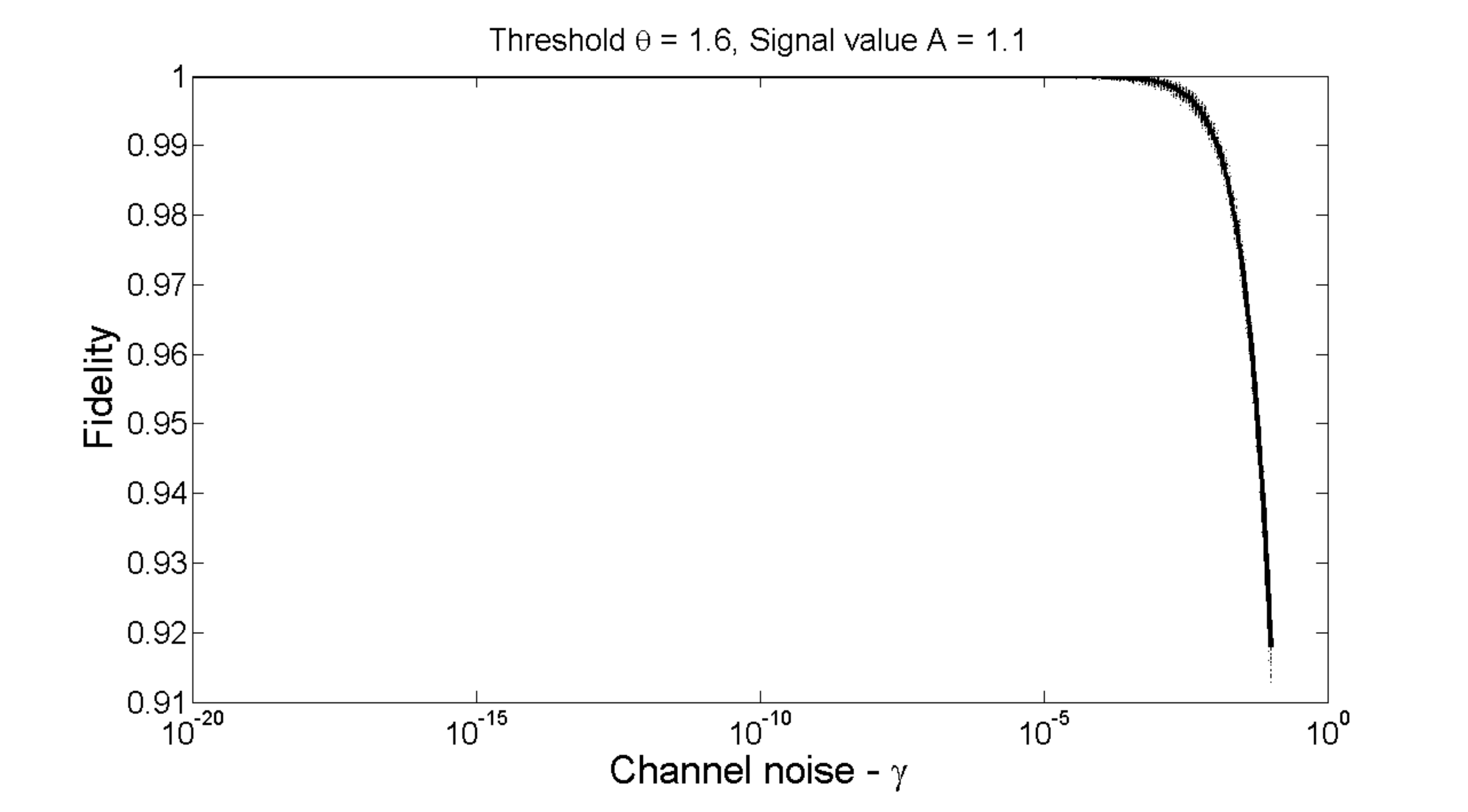}%
\\
(b)
\end{center}}}%
\end{tabular}%
\caption
{No stochastic resonance when the noise mean or location lies in the forbidden interval.
Alice possesses the state $(\left\vert0 \right\rangle+ \left\vert
1 \right\rangle)/\sqrt{2}$
and wishes to teleport it to Bob.
The graphs show
the smoothed teleportation fidelity (thick line) and min-max deviation (dotted lines) as a function of (a) the variance of
classical Gaussian noise and a function of (b) the dispersion of
classical Cauchy noise for 100 simulation runs.
Alice encodes bipolar
signals with amplitude $A=1.1$ and Bob decodes with threshold $\theta
=1.6$. Each run generated 10 000 input-output
signal pairs to estimate the fidelity of teleportation.
Graph (a) is a simulation instance of
the only-if part of Theorem~\ref{thm:finite}
with finite-variance Gaussian noise. The noise mean $\mu=.7$
and lies inside the forbidden interval $(.5,2.7)$ so that no SR occurs.
Graph (b) is a simulation instance of the only-if part of
Theorem~\ref{thm:infinite}
with infinite-variance Cauchy noise. The noise location
$a=.7$ and lies inside the forbidden interval $(.5,2.7)$ so that no SR occurs.}%
\label{fig:simulation_no_sr}%
\end{figure*}%
Theorem \ref{thm:finite} below characterizes the nonmonotone SR\ noise benefit
when the classical channel noise has finite variance. Theorem~\ref{thm:finite}
states that the modified teleportation protocol exhibits the SR\ effect if and
only if the classical noise mean obeys an interval constraint. The noise mean
must lie outside a forbidden interval that depends on Bob's detection
threshold $\theta$\ and signal value $A$. The teleportation fidelity defined
in (\ref{eq:fidelity-final}) quantifies the SR\ noise benefit.
Figure~\ref{fig:simulation}a is a simulation instance of the if-part of
Theorem~\ref{thm:finite} and Figure~\ref{fig:simulation_no_sr}a is a
simulation instance of the only-if part of Theorem~\ref{thm:finite} when the
classical channel noise is Gaussian distributed. The significance of
Theorem~\ref{thm:finite} is that it holds for any finite-variance noise
regardless of the particular density of the noise.

\begin{theorem}
\label{thm:finite}Suppose that the channel noise has finite variance
$\sigma^{2}$ and mean $\mu$. Suppose that there is some statistical dependence
between Alice's classical signal $S$ and Bob's threshold result $Y$ so that
the fidelity obeys $F>1/2$. Then the quantum teleportation system features the
nonmonotone SR effect if and only if the noise mean does not lie in the
forbidden interval: $\mu\notin\left(  \theta-A,\theta+A\right)  $. The
nonmonotone SR effect is that $F\rightarrow1/2$ as $\sigma^{2}\rightarrow0$.
\end{theorem}

\subsection{SR\ with Infinite-Variance Noise}

The uncountably infinite family of alpha-stable noise densities models many
diverse physical phenomena that include impulsive interrupts in phone lines,
underwater acoustics, low-frequency atmospheric signals, and gravitational
fluctuations \cite{book1995nikias}. The parameter $\alpha$ for the
alpha-stable noise density lies in the interval $\left(  0,2\right]  $. It
characterizes the thickness of the curve's tails: $\alpha=1$ corresponds to
the thick-tailed Cauchy random variable and $\alpha=2$ corresponds to the
familiar thin-tailed Gaussian random variable. The curve's tail thickness
increases as $\alpha$ decreases. The generalized central limit theorem states
that all and only normalized stable random variables converge in distribution
to a stable random variable \cite{book1968breiman}. The characteristic
function $\varphi\left(  \omega\right)  $ of a general alpha-stable random
variable is%
\begin{equation}
\varphi\left(  \omega\right)  =\exp\left\{  ia\omega-\gamma\left\vert
\omega\right\vert ^{\alpha}\left(  1+i\beta\text{sign}(\omega)\tan\left(
\frac{\alpha\pi}{2}\right)  \right)  \right\}  , \label{eq:alpha-stable-1}%
\end{equation}
for $\alpha\neq1$ and%
\begin{equation}
\varphi\left(  \omega\right)  =\exp\left\{  ia\omega-\gamma\left\vert
\omega\right\vert \left(  1-2i\beta\text{sign}(\omega)\ln\left(  \left\vert
\omega\right\vert \right)  /\pi\right)  \right\}  , \label{eq:alpha-stable-2}%
\end{equation}
for $\alpha=1$ where%
\begin{equation}
\text{sign}(\omega)=\left\{
\begin{array}
[c]{cc}%
1 & :\omega>0\\
0 & :\omega=0\\
-1 & :\omega<0
\end{array}
\right.  ,
\end{equation}
and $i=\sqrt{-1}$, $0<\alpha\leq2$, $-1\leq\beta\leq1$, and $\gamma>0$.
Parameter $\beta$ is a skewness parameter where $\beta=0$ gives a symmetric
density. Theorem~\ref{thm:infinite} holds for any skewness $\beta$. Parameter
$\gamma$ is a dispersion parameter similar in spirit to the variance. It
quantifies the spread or width of the alpha-stable density. Thick-tailed
alpha-stable noise may corrupt Alice's classical bipolar signals if she sends
them over an impulsive phone line or as a low-frequency signal through the atmosphere.

Theorem \ref{thm:infinite} characterizes the nonmonotone SR\ noise benefit
when the classical channel noise has an infinite-variance alpha-stable
density. Figure~\ref{fig:simulation}b is a simulation instance of the if-part
of Theorem~\ref{thm:infinite} and Figure~\ref{fig:simulation_no_sr}b is a
simulation instance of the only-if part of Theorem~\ref{thm:infinite} when the
classical channel noise is infinite-variance Cauchy distributed.
Theorem~\ref{thm:infinite} demonstrates that the SR\ noise benefit for quantum
communication is robust because it occurs even in situations when the
classical noise has infinite variance.

\begin{theorem}
\label{thm:infinite}Suppose the channel noise has an infinite-variance
alpha-stable density with dispersion $\gamma$ and location $a$. Suppose that
there is some statistical dependence between Alice's classical signal $S$ and
Bob's threshold result $Y$ so that the fidelity obeys $F>1/2$. Then the
quantum teleportation system features the nonmonotone SR effect if and only if
the noise location does not lie in the forbidden interval: $a\notin\left(
\theta-A,\theta+A\right)  $. The nonmonotone SR effect is that $F\rightarrow
1/2$ as $\gamma\rightarrow0$.
\end{theorem}

\section{Imperfect Entanglement}

\label{sec:imperfect}The entanglement shared between Alice and Bob may not
always be perfect, and it is natural to wonder whether the SR\ effect still
occurs. I briefly show that variations of the above forbidden-interval
theorems hold for the more realistic case where the shared entanglement is in
an imperfectly entangled Werner-like state \cite{PhysRevA.40.4277}. Thus, the
SR effect still occurs when the entanglement is imperfect.

Let us now suppose that Alice and Bob share the following Werner-like state as
the entanglement resource for teleportation:%
\[
\rho_{W}=F_{W}\left\vert \Phi^{+}\right\rangle \left\langle \Phi
^{+}\right\vert ^{AB}+\left(  1-F_{W}\right)  \pi^{A}\otimes\pi^{B},
\]
where $\pi$ is the maximally mixed qubit state. We can interpret the above
state as being a perfectly entangled ebit with probability $F_{W}$ and being
in a completely mixed state with probability $1-F_{W}$.

Let us consider using the above imperfectly entangled resource for the
modified teleportation protocol. Suppose that Alice and Bob perform the
modified teleportation protocol. It is straightforward to show that Bob's
resulting state is as it was before with probability $F_{W}$ and it is the
completely mixed state with probability $1-F_{W}$. Then, omitting the details,
the resulting expression for the fidelity is%
\begin{align*}
F  & =F_{W}\left(  \frac{1}{2}+\frac{P\left(  q_{X}+q_{Z}+q_{XZ}P\right)  }%
{2}\right)  +\frac{1-F_{W}}{2}\\
& =\frac{1}{2}+\frac{F_{W}P\left(  q_{X}+q_{Z}+q_{XZ}P\right)  }{2}.
\end{align*}
The above expression is similar to the one we obtained before, with the
difference that the fidelity now depends on the parameter $F_{W}$ from the
Werner-like state. Thus, the fidelity of teleportation in this
\textquotedblleft imperfect entanglement\textquotedblright\ scenario still
bears the SR\ signature because we can apply all of the above
forbidden-interval theorems.

\section{Conclusion}

The theorems for the SR\ noise benefit prove that small amounts of noise can
enhance the fidelity of quantum teleportation given the modified teleportation
protocol. The theorems lend credence to the conjecture in \cite{nn2003kosko}%
\ that an SR\ noise benefit should occur in any nonlinear system whose
input-output structure is a threshold system. The theorems show that the
SR\ effect is robust because it occurs for all finite-variance noise types and
for infinite-variance alpha-stable noise.

The theorems do not guarantee a specific performance for the teleportation
fidelity. They do not even guarantee that the teleportation fidelity exceeds
the classical limit. Figure~\ref{fig:simulation}b is an example of a failure
to exceed the classical limit of 2/3 due to impulsive Cauchy noise. The
theorems guarantee only that performance with noise exceeds performance
without noise given the satisfaction of the forbidden-interval condition.

Some may question whether the modified teleportation protocol leads to a true
\textquotedblleft quantum\textquotedblright\ stochastic resonance. It is after
all not quantum noise that affects the fidelity in this model but rather
classical noise. But several quantum effects are present in the modified
teleportation protocol such as entanglement, Bell measurements, and the
coherence of the quantum state being teleported. The interplay of quantum
effects with the noisy classical channel argues that we should categorize this
result as a \textit{classical-noise-assisted quantum stochastic resonance}.

The theorems also suggest that the SR\ noise benefit will occur in any quantum
protocol that uses feedforward classical communication with subthreshold
signals. Protocols such as entanglement purification, distillation, gate
teleportation \cite{nat1999gott}, and the Knill-Laflamme-Milburn scheme for
linear optical quantum computation \cite{nature2001klm}\ all require classical
signals. Any quantum protocol with feedforward memoryless classical
communication should exhibit the SR noise benefit when the sender transmits
subthreshold classical signals over a noisy channel and the receiver performs
threshold detection.

\begin{acknowledgments}
The author thanks Bart Kosko for useful discussions. The author thanks Todd A.
Brun, Igor Devetak, Jonathan P. Dowling, and Austin Lund for useful comments.
The author thanks Harsha Honnappa for feedback on the manuscript. The author
thanks the Hearne Institute for Theoretical Physics, the Army Research Office,
and the Disruptive Technologies Office for financial support. The author
conducted this work during a summer visit at the Hearne Institute for
Theoretical Physics in 2006.
\end{acknowledgments}

\section{Appendix:\ Proofs\label{sec:appendix}}

\begin{proof}
[Proof (Lemma \ref{lem:independent-equi})]Define states $\left\vert \Phi
^{+}\right\rangle ^{A^{\prime}A}$, $\left\vert \Phi^{-}\right\rangle
^{A^{\prime}A}$, $\left\vert \Psi^{+}\right\rangle ^{A^{\prime}A},$ and
$\left\vert \Psi^{-}\right\rangle ^{A^{\prime}A}$ as the Bell basis states:%
\begin{align}
\left\vert \Phi^{+}\right\rangle ^{A^{\prime}A}  &  \equiv\frac{\left\vert
0\right\rangle ^{A^{\prime}}\left\vert 0\right\rangle ^{A}+\left\vert
1\right\rangle ^{A^{\prime}}\left\vert 1\right\rangle ^{A}}{\sqrt{2}},\\
\left\vert \Phi^{-}\right\rangle ^{A^{\prime}A}  &  \equiv\frac{\left\vert
0\right\rangle ^{A^{\prime}}\left\vert 0\right\rangle ^{A}-\left\vert
1\right\rangle ^{A^{\prime}}\left\vert 1\right\rangle ^{A}}{\sqrt{2}},\\
\left\vert \Psi^{+}\right\rangle ^{A^{\prime}A}  &  \equiv\frac{\left\vert
0\right\rangle ^{A^{\prime}}\left\vert 1\right\rangle ^{A}+\left\vert
1\right\rangle ^{A^{\prime}}\left\vert 0\right\rangle ^{A}}{\sqrt{2}},\\
\left\vert \Psi^{-}\right\rangle ^{A^{\prime}A}  &  \equiv\frac{\left\vert
0\right\rangle ^{A^{\prime}}\left\vert 1\right\rangle ^{A}-\left\vert
1\right\rangle ^{A^{\prime}}\left\vert 0\right\rangle ^{A}}{\sqrt{2}}.
\end{align}
Make the following additional assignments:\ $\left\vert \Phi^{+}\right\rangle
^{A^{\prime}A}\equiv\left\vert \Phi_{00}\right\rangle ^{A^{\prime}A}$,
$\left\vert \Phi^{-}\right\rangle ^{A^{\prime}A}\equiv\left\vert \Phi
_{01}\right\rangle ^{A^{\prime}A}$, $\left\vert \Psi^{+}\right\rangle
^{A^{\prime}A}\equiv\left\vert \Phi_{10}\right\rangle ^{A^{\prime}A}$,
$\left\vert \Psi^{-}\right\rangle ^{A^{\prime}A}\equiv\left\vert \Phi
_{11}\right\rangle ^{A^{\prime}A}$. Define the rotated states $\left\vert
\psi_{00}\right\rangle ^{B}$, $\left\vert \psi_{01}\right\rangle ^{B}$,
$\left\vert \psi_{10}\right\rangle ^{B},$ and $\left\vert \psi_{11}%
\right\rangle ^{B}$ as follows:%
\begin{align}
\left\vert \psi_{00}\right\rangle ^{B}  &  \equiv\left\vert \psi\right\rangle
^{B}\equiv\alpha\left\vert 0\right\rangle ^{B}+\beta\left\vert 1\right\rangle
^{B},\\
\left\vert \psi_{01}\right\rangle ^{B}  &  \equiv\alpha\left\vert
0\right\rangle ^{B}-\beta\left\vert 1\right\rangle ^{B}=\hat{Z}\left\vert
\psi\right\rangle ^{B},\\
\left\vert \psi_{10}\right\rangle ^{B}  &  \equiv\alpha\left\vert
1\right\rangle ^{B}+\beta\left\vert 0\right\rangle ^{B}=\hat{X}\left\vert
\psi\right\rangle ^{B},\\
\left\vert \psi_{11}\right\rangle ^{B}  &  \equiv\alpha\left\vert
1\right\rangle ^{B}-\beta\left\vert 0\right\rangle ^{B}=\hat{X}\hat
{Z}\left\vert \psi\right\rangle ^{B}.
\end{align}
Write the joint state $\left\vert \phi\right\rangle ^{A^{\prime}AB}$ from
(\ref{eq:joint-state}) in the following form by performing a few algebraic
steps \cite{PhysRevLett.70.1895}.%
\begin{equation}
\left\vert \phi\right\rangle ^{A^{\prime}AB}=\frac{1}{2}\left[
\begin{array}
[c]{c}%
\left\vert \Phi_{00}\right\rangle ^{A^{\prime}A}\otimes\left\vert \psi
_{00}\right\rangle ^{B}+\\
\left\vert \Phi_{01}\right\rangle ^{A^{\prime}A}\otimes\left\vert \psi
_{01}\right\rangle ^{B}+\\
\left\vert \Phi_{10}\right\rangle ^{A^{\prime}A}\otimes\left\vert \psi
_{10}\right\rangle ^{B}+\\
\left\vert \Phi_{11}\right\rangle ^{A^{\prime}A}\otimes\left\vert \psi
_{11}\right\rangle ^{B}%
\end{array}
\right]  .
\end{equation}
Alice performs a measurement in the Bell basis on her two qubits $A^{\prime}$
and $A$. The joint state $\left\vert \phi\right\rangle ^{A^{\prime}AB}$
collapses to one of four states in the set $\{\left\vert \Phi_{00}%
\right\rangle ^{A^{\prime}A}\otimes\left\vert \psi_{00}\right\rangle ^{B}$,
$\left\vert \Phi_{01}\right\rangle ^{A^{\prime}A}\otimes\left\vert \psi
_{01}\right\rangle ^{B}$, $\left\vert \Phi_{10}\right\rangle ^{A^{\prime}%
A}\otimes\left\vert \psi_{10}\right\rangle ^{B}$, $\left\vert \Phi
_{11}\right\rangle ^{A^{\prime}A}\otimes\left\vert \psi_{11}\right\rangle
^{B}\}$. Alice's measurement in the Bell basis gives two classical bits
$s_{1}s_{2}$ given by the subscripts in the above equation. Suppose the two
bits $s_{1}$ and $s_{2}$ are realizations of two Bernoulli random variables
$S_{1}$ and $S_{2}$ respectively. Squaring the probability amplitudes of each
state in the superposition of quantum state $\left\vert \phi\right\rangle
^{A^{\prime}AB}$ gives an equal probability of $1/4$ for each possible state
resulting from the Bell measurement. The joint distribution of $S_{1}$ and
$S_{2}$ is as follows:%
\begin{equation}
p_{S_{1},S_{2}}\left(  s_{1},s_{2}\right)  =\frac{1}{4}\text{ \ \ \ \ }\forall
s_{1},s_{2}\in\left\{  0,1\right\}  .
\end{equation}
Thus both $S_{1}$ and $S_{2}$ are uniform random variables with the same
density. The marginal probabilities must also be uniform:%
\begin{align}
p_{S_{1}}\left(  s_{1}\right)   &  =\frac{1}{2}\text{ \ \ \ \ }\forall
s_{1}\in\left\{  0,1\right\}  ,\\
p_{S_{2}}\left(  s_{2}\right)   &  =\frac{1}{2}\text{ \ \ \ \ }\forall
s_{2}\in\left\{  0,1\right\}  .
\end{align}
$S_{1}$ and $S_{2}$ are independent random variables because the joint density
is the product of the marginals.
\end{proof}

\bigskip

\begin{proof}
[Proof (Lemma \ref{lem:fidelity})]Random variables $Y_{1}$ and $S_{1}$ and
random variables $Y_{2}$ and $S_{2}$ are independent because of the reasoning
in Section~\ref{sec:model}. Consider the joint density $p_{Y_{1},Y_{2}%
,S_{1},S_{2}}\left(  y_{1},y_{2},s_{1},s_{2}\right)  $ for Bob's two random
variables $Y_{1}$ and $Y_{2}$ and Alice's two random variables $S_{1}$ and
$S_{2}$:%
\begin{align}
&  p_{Y_{1},Y_{2},S_{1},S_{2}}\left(  y_{1},y_{2},s_{1},s_{2}\right) \\
&  =p_{Y_{1},S_{1}|Y_{2},S_{2}}\left(  y_{1},s_{1}|y_{2},s_{2}\right)
p_{Y_{2},S_{2}}\left(  y_{2},s_{2}\right) \\
&  =p_{Y_{1},S_{1}}\left(  y_{1},s_{1}\right)  p_{Y_{2},S_{2}}\left(
y_{2},s_{2}\right) \\
&  =p_{Y_{1}|S_{1}}\left(  y_{1}|s_{1}\right)  p_{S_{1}}\left(  s_{1}\right)
p_{Y_{2}|S_{2}}\left(  y_{2}|s_{2}\right)  p_{S_{2}}\left(  s_{2}\right) \\
&  =p_{Y_{1}|S_{1}}\left(  y_{1}|s_{1}\right)  \left(  \frac{1}{2}\right)
p_{Y_{2}|S_{2}}\left(  y_{2}|s_{2}\right)  \left(  \frac{1}{2}\right) \\
&  =\frac{1}{4}p_{Y|S}\left(  y_{1}|s_{1}\right)  p_{Y|S}\left(  y_{2}%
|s_{2}\right)  . \label{eq:cond-from-joint}%
\end{align}
Consider the projectors $\left\vert \psi_{y_{1}y_{2}s_{1}s_{2}}\right\rangle
\left\langle \psi_{y_{1}y_{2}s_{1}s_{2}}\right\vert $ in Bob's mixed state
$\rho_{B}$ from (\ref{eq:bobs-state}):%
\begin{align}
&  \left\vert \psi_{y_{1}y_{2}s_{1}s_{2}}\right\rangle \left\langle
\psi_{y_{1}y_{2}s_{1}s_{2}}\right\vert \\
&  =\hat{Z}^{y_{2}}\hat{X}^{y_{1}}\left\vert \psi_{s_{1}s_{2}}\right\rangle
\left\langle \psi_{s_{1}s_{2}}\right\vert \hat{X}^{y_{1}}\hat{Z}^{y_{2}}\\
&  =\hat{Z}^{y_{2}}\hat{X}^{y_{1}}\hat{X}^{s_{1}}\hat{Z}^{s_{2}}\left\vert
\psi\right\rangle \left\langle \psi\right\vert \hat{Z}^{s_{2}}\hat{X}^{s_{1}%
}\hat{X}^{y_{1}}\hat{Z}^{y_{2}}\\
&  =\hat{Z}^{y_{2}}\hat{X}^{y_{1}\oplus s_{1}}\hat{Z}^{s_{2}}\left\vert
\psi\right\rangle \left\langle \psi\right\vert \hat{Z}^{s_{2}}\hat{X}%
^{y_{1}\oplus s_{1}}\hat{Z}^{y_{2}}\\
&  =\left(
\begin{array}
[c]{c}%
\left(  -1\right)  ^{y_{1}\oplus s_{1}}\hat{X}^{y_{1}\oplus s_{1}}\hat
{Z}^{y_{2}}\hat{Z}^{s_{2}}\\
\left\vert \psi\right\rangle \left\langle \psi\right\vert \hat{Z}^{s_{2}}%
\hat{Z}^{y_{2}}\left(  -1\right)  ^{y_{1}\oplus s_{1}}\hat{X}^{y_{1}\oplus
s_{1}}%
\end{array}
\right) \\
&  =\hat{X}^{y_{1}\oplus s_{1}}\hat{Z}^{y_{2}\oplus s_{2}}\left\vert
\psi\right\rangle \left\langle \psi\right\vert \hat{Z}^{y_{2}\oplus s_{2}}%
\hat{X}^{y_{1}\oplus s_{1}}\\
&  =\left\vert \psi_{y_{1}\oplus s_{1},y_{2}\oplus s_{2}}\right\rangle
\left\langle \psi_{y_{1}\oplus s_{1},y_{2}\oplus s_{2}}\right\vert .
\label{eq:matrices-bobs-state}%
\end{align}
So Bob's mixed state $\rho_{B}$ is as follows by substituting
(\ref{eq:cond-from-joint})\ for the joint density and substituting
($\ref{eq:matrices-bobs-state}$) for the projectors $\left\vert \psi
_{y_{1}y_{2}s_{1}s_{2}}\right\rangle \left\langle \psi_{y_{1}y_{2}s_{1}s_{2}%
}\right\vert $:%
\begin{equation}
\rho_{B}=\frac{1}{4}\sum_{\substack{y_{1},y_{2},\\s_{1},s_{2}=0}}^{1}\left(
\begin{array}
[c]{c}%
p_{Y|S}\left(  y_{1}|s_{1}\right)  p_{Y|S}\left(  y_{2}|s_{2}\right) \\
\left\vert \psi_{y_{1}\oplus s_{1},y_{2}\oplus s_{2}}\right\rangle
\left\langle \psi_{y_{1}\oplus s_{1},y_{2}\oplus s_{2}}\right\vert
\end{array}
\right)  .
\end{equation}
Now use the above expression for Bob's mixed state to compute the fidelity $F$
between Alice's original state $\left\vert \psi\right\rangle ^{A^{\prime}}$
and Bob's mixed state $\rho_{B}$:%
\begin{align}
F  &  =\left\langle \psi\left\vert \rho_{B}\right\vert \psi\right\rangle \\
&  =\left\langle \psi\left\vert \frac{1}{4}\sum_{\substack{y_{1},y_{2}%
,\\s_{1},s_{2}=0}}^{1}\left(
\begin{array}
[c]{c}%
p_{Y|S}\left(  y_{1}|s_{1}\right)  p_{Y|S}\left(  y_{2}|s_{2}\right) \\
\left\vert \psi_{y_{1}\oplus s_{1},y_{2}\oplus s_{2}}\right\rangle
\left\langle \psi_{y_{1}\oplus s_{1},y_{2}\oplus s_{2}}\right\vert
\end{array}
\right)  \right\vert \psi\right\rangle \\
&  =\frac{1}{4}\sum_{\substack{y_{1},y_{2},\\s_{1},s_{2}=0}}^{1}\left(
\begin{array}
[c]{c}%
p_{Y|S}\left(  y_{1}|s_{1}\right)  p_{Y|S}\left(  y_{2}|s_{2}\right) \\
\left\vert \left\langle \psi|\psi_{y_{1}\oplus s_{1},y_{2}\oplus s_{2}%
}\right\rangle \right\vert ^{2}%
\end{array}
\right)  . \label{eq:fidelity-eq}%
\end{align}
The quantity $\left\vert \left\langle \psi|\psi_{y_{1}\oplus s_{1},y_{2}\oplus
s_{2}}\right\rangle \right\vert ^{2}$ can take one of following four values
depending on the bit values $y_{1}$, $y_{2}$, $s_{1}$, and $s_{2}$:%
\begin{align}
\left\vert \left\langle \psi\right\vert \hat{Z}\left\vert \psi\right\rangle
\right\vert ^{2}  &  =\left\vert \alpha\right\vert ^{4}-2\left\vert
\alpha\right\vert ^{2}\left\vert \beta\right\vert ^{2}+\left\vert
\beta\right\vert ^{4},\label{eq:q_prob_1}\\
\left\vert \left\langle \psi\right\vert \hat{X}\left\vert \psi\right\rangle
\right\vert ^{2}  &  =2\left(  \left\vert \beta\right\vert ^{2}\left\vert
\alpha\right\vert ^{2}+\operatorname{Re}\left\{  \beta^{2}\left(  \alpha
^{\ast}\right)  ^{2}\right\}  \right)  ,\\
\left\vert \left\langle \psi\right\vert \hat{X}\hat{Z}\left\vert
\psi\right\rangle \right\vert ^{2}  &  =2\left(  \left\vert \beta\right\vert
^{2}\left\vert \alpha\right\vert ^{2}-\operatorname{Re}\left\{  \beta
^{2}\left(  \alpha^{\ast}\right)  ^{2}\right\}  \right)  ,\\
\left\vert \left\langle \psi|\psi\right\rangle \right\vert ^{2}  &
=1=\left\vert \alpha\right\vert ^{4}+2\left\vert \alpha\right\vert
^{2}\left\vert \beta\right\vert ^{2}+\left\vert \beta\right\vert ^{4}.
\label{eq:q-prob-2}%
\end{align}
Define the nonnegative quantities $q_{Z}$, $q_{X}$, and $q_{XZ}$ as in
(\ref{eq:q_Z}-\ref{eq:q_XZ}). The quantities $q_{Z}$, $q_{X}$, and $q_{XZ}$
sum to one using (\ref{eq:q_prob_1}-\ref{eq:q-prob-2}):$\ q_{Z}+q_{X}%
+q_{XZ}=1$. Use the following shorthand for the conditional probabilities:%
\begin{align}
p_{y_{1}|s_{1}}  &  \equiv p_{Y|S}\left(  y_{1}|s_{1}\right)  ,\\
p_{y_{2}|s_{2}}  &  \equiv p_{Y|S}\left(  y_{2}|s_{2}\right)  .
\end{align}
I now prove that conditional probability differences $p_{0|0}-p_{0|1}$ and
$p_{1|1}-p_{1|0}$ are nonnegative and equal to each other. Consider the
conditional probability $p_{0|0}$:%
\begin{align}
p_{0|0}  &  =p_{Y|S}\left(  0|0\right) \\
&  =\Pr\left\{  u\left(  \left(  -1\right)  ^{S+1}A+N-\theta\right)
=0|S=0\right\} \\
&  =\Pr\left\{  u\left(  -A+N-\theta\right)  =0\right\} \\
&  =\Pr\left\{  -A+N-\theta<0\right\} \\
&  =\Pr\left\{  N<\theta+A\right\} \\
&  =\int_{-\infty}^{\theta+A}p_{N}\left(  n\right)  \ dn.
\end{align}
where $p_{N}\left(  n\right)  $ is the density of the noise $N$. The other
three conditional probabilities follow from similar reasoning:%
\begin{align}
p_{0|1}  &  =\int_{-\infty}^{\theta-A}p_{N}\left(  n\right)  \ dn,\\
p_{1|1}  &  =\int_{\theta-A}^{\infty}p_{N}\left(  n\right)  \ dn,\\
p_{1|0}  &  =\int_{\theta+A}^{\infty}p_{N}\left(  n\right)  \ dn.
\end{align}
So the conditional probability differences are equal and nonnegative because
$p_{N}\left(  n\right)  $ is nonnegative:%
\begin{align}
p_{0|0}-p_{0|1}  &  =\int_{\theta-A}^{\theta+A}p_{N}\left(  n\right)  \ dn,\\
p_{1|1}-p_{1|0}  &  =\int_{\theta-A}^{\theta+A}p_{N}\left(  n\right)  \ dn.
\end{align}
Define the nonnegative quantity $P\equiv p_{0|0}-p_{0|1}=p_{1|1}-p_{1|0}$. Let
us return to the proof of the fidelity expression. Expand the fidelity $F$
from (\ref{eq:fidelity-eq}) as follows:%
\begin{equation}
=\frac{1}{4}\left(
\begin{array}
[c]{c}%
p_{0|0}p_{0|0}+p_{0|0}p_{0|1}q_{Z}+p_{0|1}p_{0|0}q_{X}+\\
p_{0|1}p_{0|1}q_{XZ}+p_{0|0}p_{1|0}q_{Z}+p_{0|0}p_{1|1}+\\
p_{0|1}p_{1|0}q_{XZ}+p_{0|1}p_{1|1}q_{X}+p_{1|0}p_{0|0}q_{X}+\\
p_{1|0}p_{0|1}q_{XZ}+p_{1|1}p_{0|0}+p_{1|1}p_{0|1}q_{Z}+\\
p_{1|0}p_{1|0}q_{XZ}+p_{1|0}p_{1|1}q_{X}+p_{1|1}p_{1|0}q_{Z}+\\
p_{1|1}p_{1|1}%
\end{array}
\right)
\end{equation}%
\begin{equation}
=\frac{1}{4}\left(
\begin{array}
[c]{c}%
p_{0|0}p_{0|0}+p_{0|0}p_{1|1}+\\
p_{1|1}p_{0|0}+p_{1|1}p_{1|1}+\\
\left[
\begin{array}
[c]{c}%
p_{0|0}p_{1|0}+p_{0|0}p_{0|1}+\\
p_{1|1}p_{0|1}+p_{1|1}p_{1|0}%
\end{array}
\right]  q_{Z}+\\
\left[
\begin{array}
[c]{c}%
p_{0|1}p_{0|0}+p_{0|1}p_{1|1}+\\
p_{1|0}p_{0|0}+p_{1|0}p_{1|1}%
\end{array}
\right]  q_{X}+\\
\left[
\begin{array}
[c]{c}%
p_{0|1}p_{0|1}+p_{0|1}p_{1|0}+\\
p_{1|0}p_{0|1}+p_{1|0}p_{1|0}%
\end{array}
\right]  q_{XZ}%
\end{array}
\right)
\end{equation}%
\begin{equation}
=\frac{1}{4}\left(
\begin{array}
[c]{c}%
p_{0|0}p_{0|0}+p_{0|0}p_{1|1}+\\
p_{1|1}p_{0|0}+p_{1|1}p_{1|1}+\\
\left[
\begin{array}
[c]{c}%
p_{0|0}p_{1|0}+p_{0|0}p_{0|1}+\\
p_{1|1}p_{0|1}+p_{1|1}p_{1|0}%
\end{array}
\right]  \left(  q_{Z}+q_{X}\right)  +\\
\left[
\begin{array}
[c]{c}%
p_{0|1}p_{0|1}+p_{0|1}p_{1|0}+\\
p_{1|0}p_{0|1}+p_{1|0}p_{1|0}%
\end{array}
\right]  q_{XZ}%
\end{array}
\right)
\end{equation}%
\begin{equation}
=\frac{1}{4}\left(
\begin{array}
[c]{c}%
p_{0|0}p_{0|0}+p_{0|0}p_{1|1}+\\
p_{1|1}p_{0|0}+p_{1|1}p_{1|1}+\\
\left[
\begin{array}
[c]{c}%
p_{0|0}p_{1|0}+p_{0|0}p_{0|1}+\\
p_{1|1}p_{0|1}+p_{1|1}p_{1|0}%
\end{array}
\right]  \left(  1-q_{XZ}\right)  +\\
\left[
\begin{array}
[c]{c}%
p_{0|1}p_{0|1}+p_{0|1}p_{1|0}+\\
p_{1|0}p_{0|1}+p_{1|0}p_{1|0}%
\end{array}
\right]  q_{XZ}%
\end{array}
\right)
\end{equation}%
\begin{equation}
=\frac{1}{4}\left(
\begin{array}
[c]{c}%
p_{0|0}+p_{0|0}+\\
p_{1|1}+p_{1|1}+\\
\left[
\begin{array}
[c]{c}%
Pp_{0|1}+Pp_{1|0}+\\
Pp_{0|1}+Pp_{1|0}%
\end{array}
\right]  \left(  -q_{XZ}\right)
\end{array}
\right)
\end{equation}%
\begin{equation}
=\frac{1}{4}\left(
\begin{array}
[c]{c}%
2+2P\\
+\left[  2P\left(  p_{0|1}+p_{1|0}\right)  \right]  \left(  -q_{XZ}\right)
\end{array}
\right)
\end{equation}%
\begin{align}
&  =\left(  2+2P\left(  1-q_{XZ}\left(  p_{0|1}+p_{1|0}\right)  \right)
\right)  /4\\
&  =\left(  1+P\left(  1-q_{XZ}\left(  1-p_{1|1}+p_{1|0}\right)  \right)
\right)  /2\\
&  =\left(  1+P\left(  1-q_{XZ}+q_{XZ}P\right)  \right)  /2\\
&  =\frac{1}{2}+\frac{P\left(  q_{X}+q_{Z}+q_{XZ}P\right)  }{2}.
\end{align}
The quantities $q_{X}+q_{Z},q_{XZ},P$ in the above expression are nonnegative.
\end{proof}

\bigskip

\begin{proof}
[Proof (Corollary \ref{lem:minimum-fidelity})]First characterize the
relationship between random variables $S$ and $Y$ and parameter $P$. Then
translate this relationship to the fidelity $F$ by using
(\ref{eq:fidelity-final}). Expand the probability $p_{Y}\left(  y\right)  $
using the law of total probability:%
\begin{align}
p_{Y}\left(  y\right)   &  =p_{Y|S}\left(  y|0\right)  p_{S}\left(  0\right)
+p_{Y|S}\left(  y|1\right)  p_{S}\left(  1\right) \\
&  =p_{Y|S}\left(  y|0\right)  p_{S}\left(  0\right)  +p_{Y|S}\left(
y|1\right)  \left(  1-p_{S}\left(  0\right)  \right) \\
&  =\left(  p_{Y|S}\left(  y|0\right)  -p_{Y|S}\left(  y|1\right)  \right)
p_{S}\left(  0\right)  +p_{Y|S}\left(  y|1\right)  .
\end{align}
Consider when $y=0$:%
\begin{align}
p_{Y}\left(  0\right)   &  =\left(  p_{Y|S}\left(  0|0\right)  -p_{Y|S}\left(
0|1\right)  \right)  p_{S}\left(  0\right)  +p_{Y|S}\left(  0|1\right) \\
&  =P\ p_{S}\left(  0\right)  +p_{Y|S}\left(  0|1\right)  .
\end{align}
The probability $p_{Y}\left(  0\right)  =p_{Y|S}\left(  0|1\right)  $ when
parameter $P=0$. Expand the probability $p_{Y}\left(  y\right)  $ in a similar
manner so that%
\begin{equation}
p_{Y}\left(  y\right)  =\left(  p_{Y|S}\left(  y|1\right)  -p_{Y|S}\left(
y|0\right)  \right)  p_{S}\left(  1\right)  +p_{Y|S}\left(  y|0\right)  .
\end{equation}
Consider when $y=1$:%
\begin{align}
p_{Y}\left(  1\right)   &  =\left(  p_{Y|S}\left(  1|1\right)  -p_{Y|S}\left(
1|0\right)  \right)  p_{S}\left(  1\right)  +p_{Y|S}\left(  1|0\right) \\
&  P\ p_{S}\left(  1\right)  +p_{Y|S}\left(  1|0\right)  .
\end{align}
The probability $p_{Y}\left(  1\right)  =p_{Y|S}\left(  1|0\right)  $ when
parameter $P=0$. Random variables $Y$ and $S$ are independent if and only if
parameter $P=0$ because the probabilities $p_{Y}\left(  0\right)  $ and
$p_{Y}\left(  1\right)  $ are equal to probabilities conditioned on $S$.
Detection is perfect if and only if the conditional probabilities
$p_{Y|S}\left(  0|0\right)  =p_{Y|S}\left(  1|1\right)  =1$. Suppose $P=1$.
Then%
\begin{align}
1  &  =p_{Y|S}\left(  0|0\right)  -p_{Y|S}\left(  0|1\right) \\
&  =p_{Y|S}\left(  0|0\right)  -\left(  1-p_{Y|S}\left(  1|1\right)  \right)
\\
&  =p_{Y|S}\left(  0|0\right)  -1+p_{Y|S}\left(  1|1\right) \\
\Leftrightarrow2  &  =p_{Y|S}\left(  0|0\right)  +p_{Y|S}\left(  1|1\right)  .
\end{align}
Both $p_{Y|S}\left(  0|0\right)  =p_{Y|S}\left(  1|1\right)  =1$ because they
are probabilities and neither $p_{Y|S}\left(  0|0\right)  $ nor $p_{Y|S}%
\left(  1|1\right)  $ can be greater than one. So detection is perfect if and
only if $P=1$. Parameter $P$ varies between zero and one. The fidelity $F$
becomes minimum at $1/2$ if $P$ vanishes by using (\ref{eq:fidelity-final}). A
nonzero value of $P$ corresponds to statistical dependence of random variables
$S$ and $Y$ and gives a fidelity $F>1/2$. Perfect detection gives $P=1$ and
gives a perfect fidelity $F=1$ because $q_{X}$, $q_{Z}$, and $q_{XZ}$ are
nonnegative and sum to one using (\ref{eq:q_prob_1}-\ref{eq:q-prob-2}).
\end{proof}

\bigskip

\begin{proof}
[Proof (Theorem \ref{thm:finite})]Suppose the noise mean $\mu$ is not in the
forbidden interval: $\mu\notin\left(  \theta-A,\theta+A\right)  $. Then I
prove that $P$ vanishes when the finite variance $\sigma^{2}\rightarrow0$. The
fidelity $F$ approaches its minimum of $1/2$ when $P\rightarrow0$. Thus the
fidelity $F$ rises from its minimum at $1/2$ as the channel adds some noise.
The \textquotedblleft what goes down must come up\textquotedblright\ proof
strategy is the same as the earlier forbidden-interval theorem proofs
\cite{nn2003kosko,pre2004kosko}. I include the full proof for completeness. I
first prove the sufficient condition. Ignore the zero-measure case when
$\mu=\theta+A$ or $\mu=\theta-A$. Suppose the noise mean $\mu>\theta+A$. Pick
$\varepsilon=\left(  \mu-\theta-A\right)  /2>0$ so that $\theta+A+\varepsilon
=\mu-\varepsilon$. Consider parameter $P$:%
\begin{align}
P  &  =\int_{\theta-A}^{\theta+A}p_{N}\left(  n\right)  \ dn\\
&  \leq\int_{-\infty}^{\theta+A}p_{N}\left(  n\right)  \ dn\\
&  \leq\int_{-\infty}^{\theta+A+\varepsilon}p_{N}\left(  n\right)  \ dn\\
&  =\int_{-\infty}^{\mu-\varepsilon}p_{N}\left(  n\right)  \ dn\\
&  =\Pr\left\{  N<\mu-\varepsilon\right\} \\
&  =\Pr\left\{  N-\mu<-\varepsilon\right\} \\
&  \leq\Pr\left\{  \left\vert N-\mu\right\vert >\varepsilon\right\} \\
&  \leq\frac{\sigma^{2}}{\varepsilon^{2}}\rightarrow0\text{ as }\sigma
^{2}\rightarrow0.
\end{align}
Suppose the noise mean $\mu<\theta-A$. Pick $\varepsilon=\left(  \theta
-A-\mu\right)  /2>0$ so that $\theta-A-\varepsilon=\mu+\varepsilon$. Consider
parameter $P$:%
\begin{align}
P  &  =\int_{\theta-A}^{\theta+A}p_{N}\left(  n\right)  \ dn\\
&  \leq\int_{\theta-A}^{\infty}p_{N}\left(  n\right)  \ dn\\
&  \leq\int_{\theta-A-\varepsilon}^{\infty}p_{N}\left(  n\right)  \ dn\\
&  =\int_{\mu+\varepsilon}^{\infty}p_{N}\left(  n\right)  \ dn\\
&  =\Pr\left\{  N>\mu+\varepsilon\right\} \\
&  =\Pr\left\{  N-\mu>\varepsilon\right\} \\
&  \leq\Pr\left\{  \left\vert N-\mu\right\vert >\varepsilon\right\} \\
&  \leq\frac{\sigma^{2}}{\varepsilon^{2}}\rightarrow0\text{ as }\sigma
^{2}\rightarrow0.
\end{align}
I now prove the forbidden-interval condition is necessary for the SR noise
benefit. Suppose the noise mean $\mu$ is in the forbidden interval: $\mu
\in\left(  \theta-A,\theta+A\right)  $. Then I prove that parameter
$P\rightarrow1$ as $\sigma^{2}\rightarrow0$ and thus the fidelity
$F\rightarrow1$ by Corollary~\ref{lem:minimum-fidelity}. So the nonomonotone
SR\ noise benefit does not occur as the noise variance vanishes. Parameter
$P\rightarrow1$ if and only if the conditional probabilities $p_{Y|S}\left(
0|0\right)  \rightarrow1$ and $p_{Y|S}\left(  1|1\right)  \rightarrow1$.
Consider the conditional probability $p_{Y|S}\left(  0|0\right)  $. Pick
$\varepsilon=\left(  \theta+A-\mu\right)  /2$. Then $\theta+A-\varepsilon
=\mu+\varepsilon$.%
\begin{align}
p_{Y|S}\left(  0|0\right)   &  =\int_{-\infty}^{\theta+A}p_{N}\left(
n\right)  \ dn\\
&  \geq\int_{-\infty}^{\theta+A-\varepsilon}p_{N}\left(  n\right)  \ dn\\
&  =\int_{-\infty}^{\mu+\varepsilon}p_{N}\left(  n\right)  \ dn\\
&  =\Pr\left\{  N<\mu+\varepsilon\right\} \\
&  =\Pr\left\{  N-\mu<\varepsilon\right\} \\
&  =1-\Pr\left\{  N-\mu\geq\varepsilon\right\} \\
&  \geq1-\Pr\left\{  \left\vert N-\mu\right\vert \geq\varepsilon\right\} \\
&  \geq1-\frac{\sigma^{2}}{\varepsilon^{2}}\rightarrow1\text{ as }\sigma
^{2}\rightarrow0.
\end{align}
Consider the conditional probability $p_{Y|S}\left(  1|1\right)  $. Pick
$\varepsilon=\left(  \mu-\theta+A\right)  /2$. Then $\theta-A+\varepsilon
=\mu-\varepsilon$ so that%
\begin{align}
p_{Y|S}\left(  1|1\right)   &  =\int_{\theta-A}^{\infty}p_{N}\left(  n\right)
\ dn\\
&  \geq\int_{\theta-A+\varepsilon}^{\infty}p_{N}\left(  n\right)  \ dn\\
&  =\int_{\mu-\varepsilon}^{\infty}p_{N}\left(  n\right)  \ dn\\
&  =\Pr\left\{  N>\mu-\varepsilon\right\} \\
&  =\Pr\left\{  N-\mu>-\varepsilon\right\} \\
&  =1-\Pr\left\{  N-\mu\leq-\varepsilon\right\} \\
&  \geq1-\Pr\left\{  \left\vert N-\mu\right\vert \geq\varepsilon\right\} \\
&  \geq1-\frac{\sigma^{2}}{\varepsilon^{2}}\rightarrow1\text{ as }\sigma
^{2}\rightarrow0.
\end{align}
So parameter $P\rightarrow1$ because the conditional probabilities
$p_{Y|S}\left(  0|0\right)  \rightarrow1$ and $p_{Y|S}\left(  1|1\right)
\rightarrow1$. The fidelity $F\rightarrow1$ as the noise vanishes and the
nonmonotone SR noise benefit does not occur.
\end{proof}

\bigskip

\begin{proof}
[Proof (Theorem \ref{thm:infinite})]Suppose the noise location $a$ is not in
the forbidden interval: $a\notin\left(  \theta-A,\theta+A\right)  $. Then I
prove that $P$ vanishes when the dispersion $\gamma\rightarrow0$. The fidelity
$F$ approaches its minimum of $1/2$ when $P\rightarrow0$. Thus the fidelity
$F$ rises from its minimum at $1/2$ as the classical channel adds some noise.
The alpha-stable\ proof strategy is the same as the earlier alpha-stable
forbidden-interval theorem proofs \cite{nn2003kosko,pre2004kosko}. I include
the full proof for completeness. The proof for the alpha-stable case is simple
because the characteristic function in (\ref{eq:alpha-stable-1}) and
(\ref{eq:alpha-stable-2}) approaches the following as the dispersion vanishes:%
\begin{equation}
\lim_{\gamma\rightarrow0}\varphi\left(  \omega\right)  =\exp\left(
ia\omega\right)  .
\end{equation}
The inverse Fourier transform of the characteristic function gives the
limiting density as a translated delta function:\ $\lim_{\gamma\rightarrow
0}p_{N}\left(  n\right)  =\delta\left(  n-a\right)  $. I first prove the
sufficient condition. Ignore the zero-measure case when $\mu=\theta+A$ or
$\mu=\theta-A$. Consider parameter $P$:%
\begin{align*}
\lim_{\gamma\rightarrow0}P  &  =\lim_{\gamma\rightarrow0}\int_{\theta
-A}^{\theta+A}p_{N}\left(  n\right)  \ dn\\
&  =\int_{\theta-A}^{\theta+A}\delta\left(  n-a\right)  \ dn=0.
\end{align*}
So the fidelity $F$ approaches its minimum at $1/2$ as the channel noise
dispersion $\gamma$ vanishes. I now prove the forbidden-interval condition is
necessary for the nonmonotone SR noise benefit. Suppose the noise location $a$
is in the forbidden interval: $a\in\left(  \theta-A,\theta+A\right)  $. Then I
prove that parameter $P\rightarrow1$ as $\gamma\rightarrow0$ and thus the
fidelity $F\rightarrow1$ by Corollary~\ref{lem:minimum-fidelity}. So the
nonomonotone SR\ effect does not occur as the noise dispersion vanishes.
Parameter $P\rightarrow1$ if and only if the conditional probabilities
$p_{Y|S}\left(  0|0\right)  \rightarrow1$ and $p_{Y|S}\left(  1|1\right)
\rightarrow1$. Consider the conditional probability $p_{Y|S}\left(
0|0\right)  $:%
\begin{align*}
\lim_{\gamma\rightarrow0}p_{Y|S}\left(  0|0\right)   &  =\lim_{\gamma
\rightarrow0}\int_{-\infty}^{\theta+A}p_{N}\left(  n\right)  \ dn\\
&  =\int_{-\infty}^{\theta+A}\delta\left(  n-a\right)  \ dn=1.
\end{align*}
Consider the conditional probability $p_{Y|S}\left(  1|1\right)  $:%
\begin{align*}
\lim_{\gamma\rightarrow0}p_{Y|S}\left(  1|1\right)   &  =\lim_{\gamma
\rightarrow0}\int_{\theta-A}^{\infty}p_{N}\left(  n\right)  \ dn\\
&  =\int_{\theta-A}^{\infty}\delta\left(  n-a\right)  \ dn=1.
\end{align*}
So parameter $P\rightarrow1$ because the conditional probabilities
$p_{Y|S}\left(  0|0\right)  \rightarrow1$ and $p_{Y|S}\left(  1|1\right)
\rightarrow1$. The fidelity $F\rightarrow1$ as the noise vanishes and the
nonmonotone SR noise benefit does not occur.
\end{proof}

\bibliographystyle{apsrev}
\bibliography{pra-sr-tele}

\end{document}